\documentclass{iau}

\usepackage{amsmath}
\usepackage{graphicx}
\usepackage{multirow}
\usepackage{journals}
\usepackage{orcidlink}

\newcommand\msun{\ M_\odot}
\newcommand\heii{He II $\lambda 1640$}
\newcommand\hb{H$\beta$}

\defcitealias{Gotberg2019}{G19}

\begin{document}

\lefttitle{Hovis-Afflerbach et al.}
\righttitle{Testing Novel Models of Binary Populations across Cosmic Time}

\jnlPage{1}{7}
\volno{402}
\jnlDoiYr{2025}
\doival{}

\aopheadtitle{Proceedings IAU Symposium}
\editors{A. Wofford,  N. St-Louis, M. Garcia \&  S. Simón-Díaz, eds.}

\title{Finding He II: Testing Novel Models of Binary Populations across Cosmic Time}

\author{Beryl Hovis-Afflerbach\orcidlink{0000-0002-9967-2725}$^{1,2}$, Allison L. Strom\orcidlink{0000-0001-6369-1636}$^{1,2}$, Alberto Saldana-Lopez\orcidlink{0000-0001-8419-3062}$^3$, Sophia R. Flury\orcidlink{0000-0002-0159-2613}$^4$}
\affiliation{
$^1$Department of Physics and Astronomy, Northwestern University,\\2145 Sheridan Rd, Evanston, IL 60208, USA \\ 
email: {\tt \url{beryl@u.northwestern.edu}} \\ 
$^2$Center for Interdisciplinary Exploration and Research in Astrophysics (CIERA), Northwestern University, 1800 Sherman Ave, Evanston, IL 60201, USA \\ 
$^3$Department of Astronomy, Oskar Klein Centre, Stockholm University, SE-106 91 Stockholm, Sweden \\ 
$^4$Institute for Astronomy, University of Edinburgh, Royal Observatory, Edinburgh EH9 3HJ, UK
}

\begin{abstract}
Our understanding of massive stars remains incomplete. Many high-$z$ galaxies and nearby analogs exhibit strong He II emission, indicating an abundance of photons with energies $>54.4$ eV that standard single-star population models cannot explain. Recent studies show that binary evolution and non-solar abundance patterns are required to explain the distinct spectra of high-$z$ galaxies observed by JWST. However, treatments of these properties vary drastically between models. We present the first results from a comparison of models with different treatments of binaries, including BPASS and novel stripped star models, with rest-UV-optical spectra of high-$z$ galaxies' local analogs. This type of investigation can provide insights into which aspects of binary evolution are important to reproduce observations and identify priorities in ongoing efforts to improve models. By constraining the properties of massive stars at high redshift, we can learn about the processes at play in high-$z$ galaxies and massive star evolution more broadly.
\end{abstract}


\maketitle

\section{Introduction}\label{sec:intro}

Metallicity is one of the most significant factors in determining how a star evolves and shapes its environment.
It is therefore crucial to understand how the metal-poor stars that populate the high-redshift universe differ from those with Solar metallicity, in order to understand how those stars, and their host galaxies, evolved.
Ideally, it would be possible to test models of metal-poor massive stars with observations of individual stars from resolved populations in the nearby Universe, but due to many factors including shorter lifetimes and the scarcity of nearby low-metallicity environments, the population of such stars nearby is very small.
Studying metal-poor massive stars therefore requires comparisons with high-redshift star-forming galaxies and their local analogs.
In these galaxies, the ultraviolet (UV, $\sim 100-3000$ \AA) part of the spectrum should be dominated by massive stars.

Approximately $1/3$ of O stars are expected to have their hydrogen envelopes stripped by interaction with a binary companion \citep{Sana2012}.
The resulting hot, intermediate-mass ``stripped stars'' have been predicted for many years, but the first sample of 16 was only recently observed in the Magellanic Clouds \citep{Drout2023, Gotberg2023}.
Stripped stars are UV-bright, with the most massive potentially having spectra similar to those of Wolf-Rayet stars, and have long been considered as a potential source of the hard ionizing galaxy spectra that are ubiquitous at high redshift.
In optical wavelengths, however, they are often obscured by a binary companion, which likely explains why they evaded detection for so long.

It is possible that their elusiveness is not only the product of observational bias but also a reflection of true scarcity.
Studies using a variety of stellar evolution codes have predicted that low-metallicity massive stars should not expand until very late in their lifetimes, delaying binary interaction and therefore envelope stripping \citep[e.g.,][]{Klencki2020, Farrell2022}.
The resulting stripped stars would be very short-lived or perhaps only partially-stripped.
\cite{Hovis-Afflerbach2025} used a binary population synthesis model initially developed by \citet[][hereafter \citetalias{Gotberg2019}]{Gotberg2019} to model the mass distribution of stripped stars at different metallicities and found that this late expansion results in a ``helium star desert'' where massive ($M_{\rm strip} \sim 12-30 \msun$) stripped stars are missing at low metallicity ($Z \lesssim 0.002$).

\begin{figure}
\centering
\includegraphics[width=0.48\linewidth]{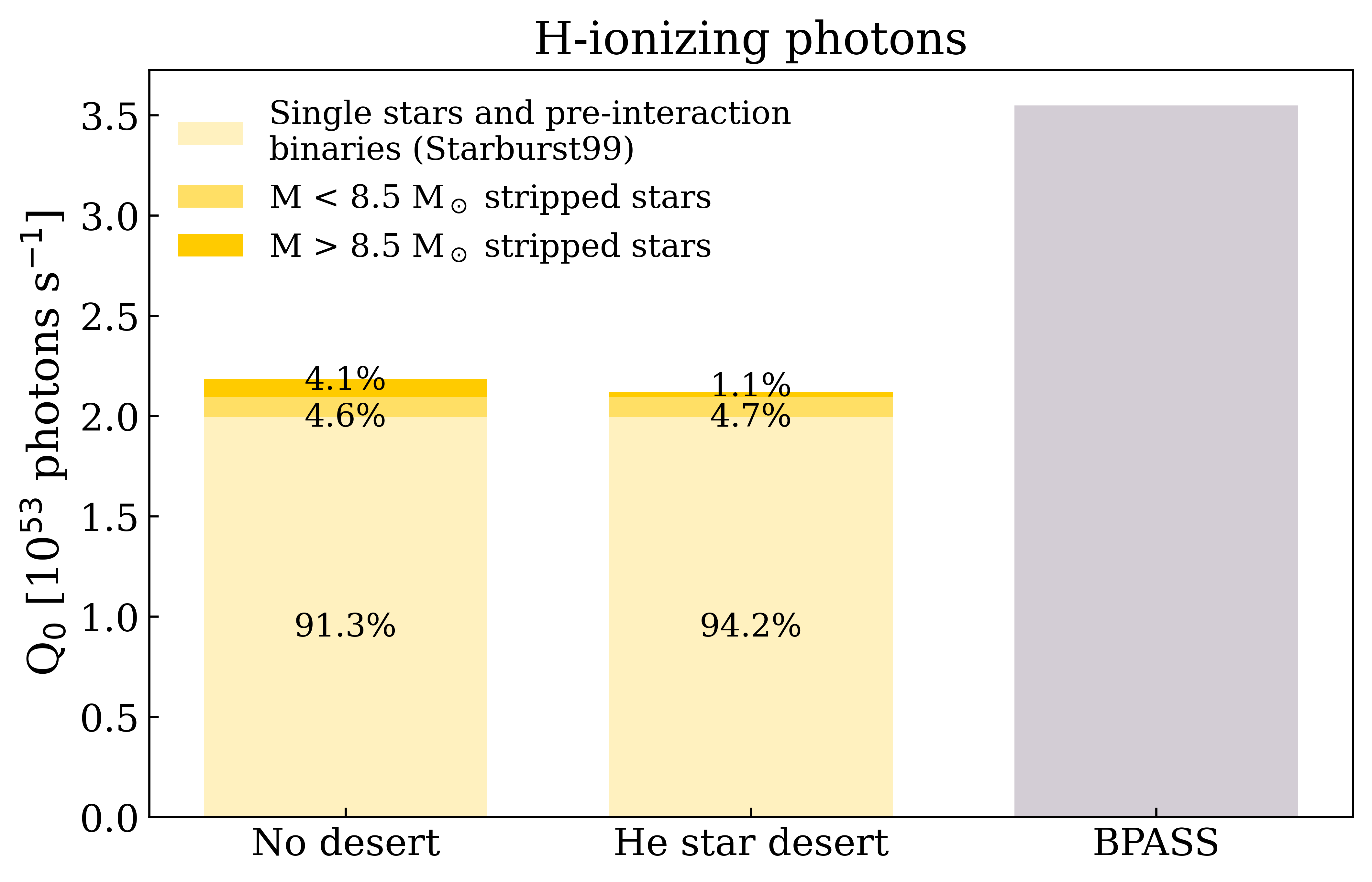}
\includegraphics[width=0.48\linewidth]{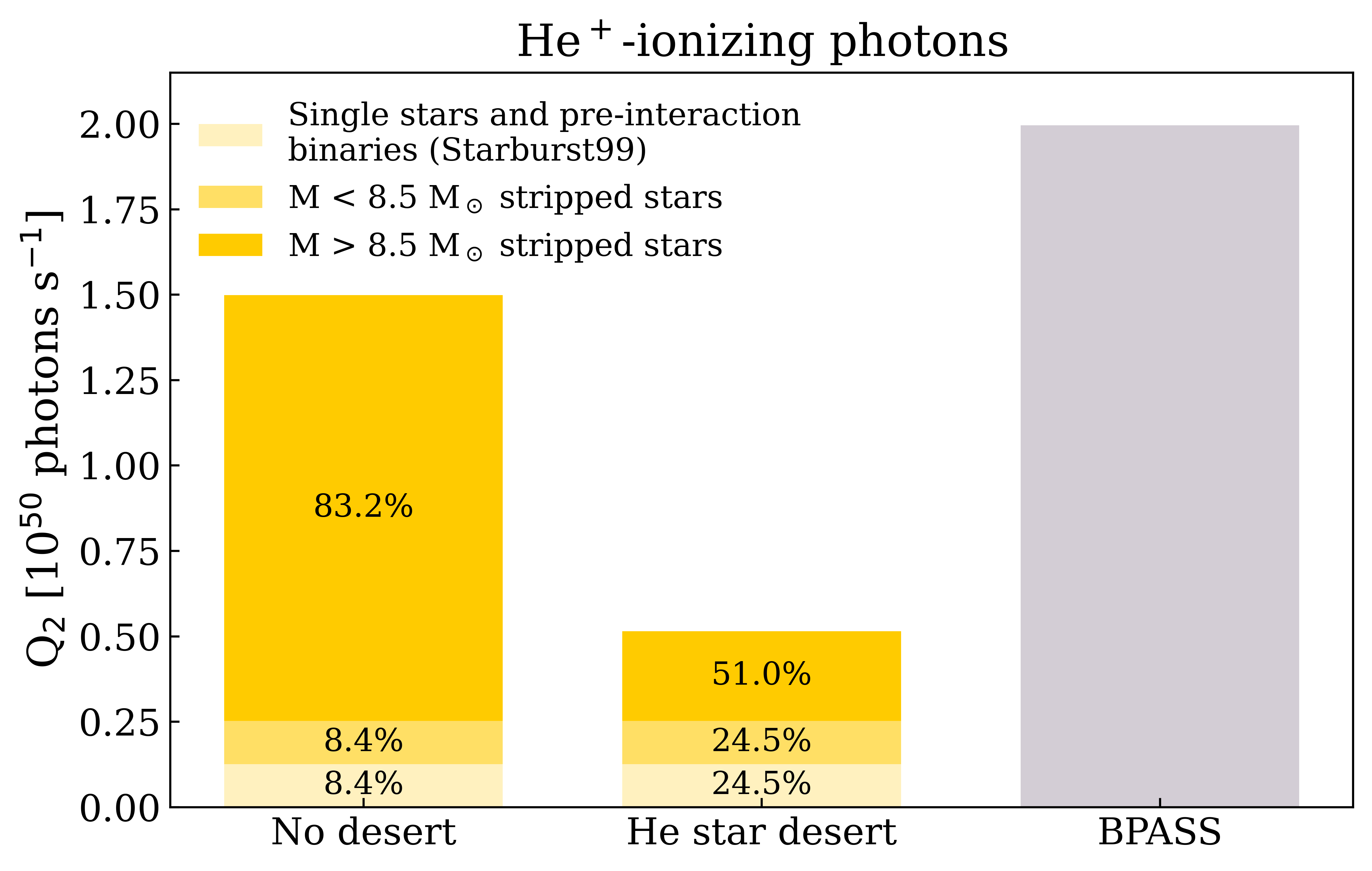}
\caption{Estimates for hydrogen- ($Q_0$, left) and helium-ionizing ($Q_2$, right) emission rates from \cite{Hovis-Afflerbach2025}. The total rates are shown as a sum of rates for single stars (light yellow), low-mass stripped stars ($M < 8.5\ M_\odot$, medium yellow), and higher-mass stripped stars ($M > 8.5\ M_\odot$, dark yellow). Stripped stars dominate $Q_2$, resulting in a significantly higher rate if there is no helium star desert (left column) than if there is (middle column).}
\label{fig:Q}
\end{figure}

Using these distributions, \cite{Hovis-Afflerbach2025} predict the impact of the helium star desert on ionizing photon production rates. Figure \ref{fig:Q} \citep[][Figure 8]{Hovis-Afflerbach2025} shows that while stripped stars have a non-negligible impact on hydrogen-ionizing photons ($Q_0$, left), they dominate the production of helium-ionizing photons ($Q_2$, right), and the $Q_2$ prediction is therefore significantly impacted by the helium star desert. The \heii \ emission line and its weaker optical counterpart are therefore excellent probes of the helium star desert and of stripped stars in general.

More efforts are being made to identify and study nearby stripped stars. \cite{Ludwig2025} present over 800 more candidate stripped stars in the Magellanic Clouds, and finding stripped stars is one of the goals of the Ultraviolet Explorer \citep[UVEX,][]{Kulkarni2021} mission, but it will be years before a sufficient sample of nearby stripped stars are detected to test for the helium star desert.
In the meantime, high-$z$, star-forming galaxies and their local analogs provide a unique testbed to study stripped stars and binary evolution at low metallicity.

Here, we compare three different population and spectral synthesis models with observations of local analog galaxies to test their ability to reproduce these galaxies' hard ionizing spectra.
This is the first observational test of the \citetalias{Gotberg2019} stripped star models.

\section{Models of massive star populations}\label{sec:models}

We compare three sets of models. 
For each, we use a grid of spectra for instantaneous $10^6 \msun$ starbursts with ages 1, 2, 3, 4, 5, 8, 10, 15, 20 and 40 Myr and metallicities of approximately $Z = 0.014,\ 0.006,\ 0.002$, and $0.001$.

Starburst99 \citep{Leitherer1999} is one of the most widely-used population synthesis codes. It does not include binaries, so we use it as a control with which to compare the two binary population models.
We use the same Starburst99 models described in Section 2.3 of \citetalias{Gotberg2019}, with stellar metallicities $Z = 0.014,\ 0.008,\ 0.002$, and $0.001$.

We use Binary Population And Spectral Synthesis v2.3 \citep[BPASS,][]{Eldridge2017} models with Solar-scaled metal abundances and stellar metallicities $Z = 0.014,\ 0.006,\ 0.002$, and $0.001$.
Because stars at higher redshifts are expected to be $\alpha$-enhanced \citep[e.g.,][]{Steidel2016}, a treatment including $\alpha$-enhancement would be more accurate. However, our other models use Solar-scaled abundances, and this version of BPASS focuses on $\alpha$-enhancement only in low-mass stars.
As model spectra including $\alpha$-enhanced massive stars become available \citep[e.g.,][]{Park2024, Byrne2025}, more realistic tests will be possible.

Finally, we use the same \citetalias{Gotberg2019} stripped star population models used by \cite{Hovis-Afflerbach2025} to predict the helium star desert.
These include only stripped stars, so we add them to the single star population spectra from Starburst99.
The contribution of stripped stars to the ionizing spectrum is significant only after 10 Myr \citepalias{Gotberg2019}, so the UV spectra of younger bursts can be approximated using purely the single stars.
We take the same approach as \citetalias{Gotberg2019} and add stripped star models with $Z = 0.014,\ 0.006,\ 0.002$, and $0.0002$ to Starburst99 models with the closest metallicities, $Z = 0.014,\ 0.008,\ 0.002$, and $0.001$, respectively.

\section{Methods}\label{sec:methods}

We compare the model spectra with CLASSY \citep{Berg2022}, a sample of nearby star-forming galaxies with similar properties as seen at high redshift.
We include the 19 galaxies with \heii \ emission detected at a signal-to-noise ratio $> 3$ \citep{Mingozzi2022}.

To simulate the nebular continuum and emission lines, we process all three grids of models, each spanning a range of stellar metallicities and ages, through Cloudy 23.00 \citep{Chatzikos2023} using identical input parameters, so any differences should arise from only differences in the models themselves.
For each model, the gas-phase metallicity is set equal to the stellar metallicity of the input spectrum, using Solar-scaled abundances (however, see Section \ref{sec:models} for notes on $\alpha$-enhancement).
We assume spherical geometry and an ionization parameter of $\log U = -2.5$.
We use an isobaric density profile with \texttt{constant pressure 6.0} and a hydrogen density at the illuminated face of $n_{\rm H} = 10^2 \ {\rm cm}^{-3}$.
The calculation stops when the hydrogen ionization fraction drops to $10^{-2}$, and each model is iterated to convergence.

Once we process each model through Cloudy, we add the resulting nebular continuum to the incident spectrum.
We then use the FiCUS code\footnote{FItting the stellar Continuum of Uv Spectra, \url{https://github.com/asalda/FiCUS}} \citep{Saldana-Lopez2023} to fit the UV continuum ($1120-1760$ \AA) of each observed galaxy.
We do so independently with each of the three grids of models, which have been implemented in FiCUS as \texttt{ssp\_models} and can be used by choosing \texttt{sb99}, \texttt{bpass}, or \texttt{sb99stripped}.

\section{Comparison of model predictions with observations}\label{sec:comparison}

\subsection{Properties derived from the nebular continuum}\label{subsec:continuum}

\begin{figure}
    \centering
    \includegraphics[width=0.8\linewidth]{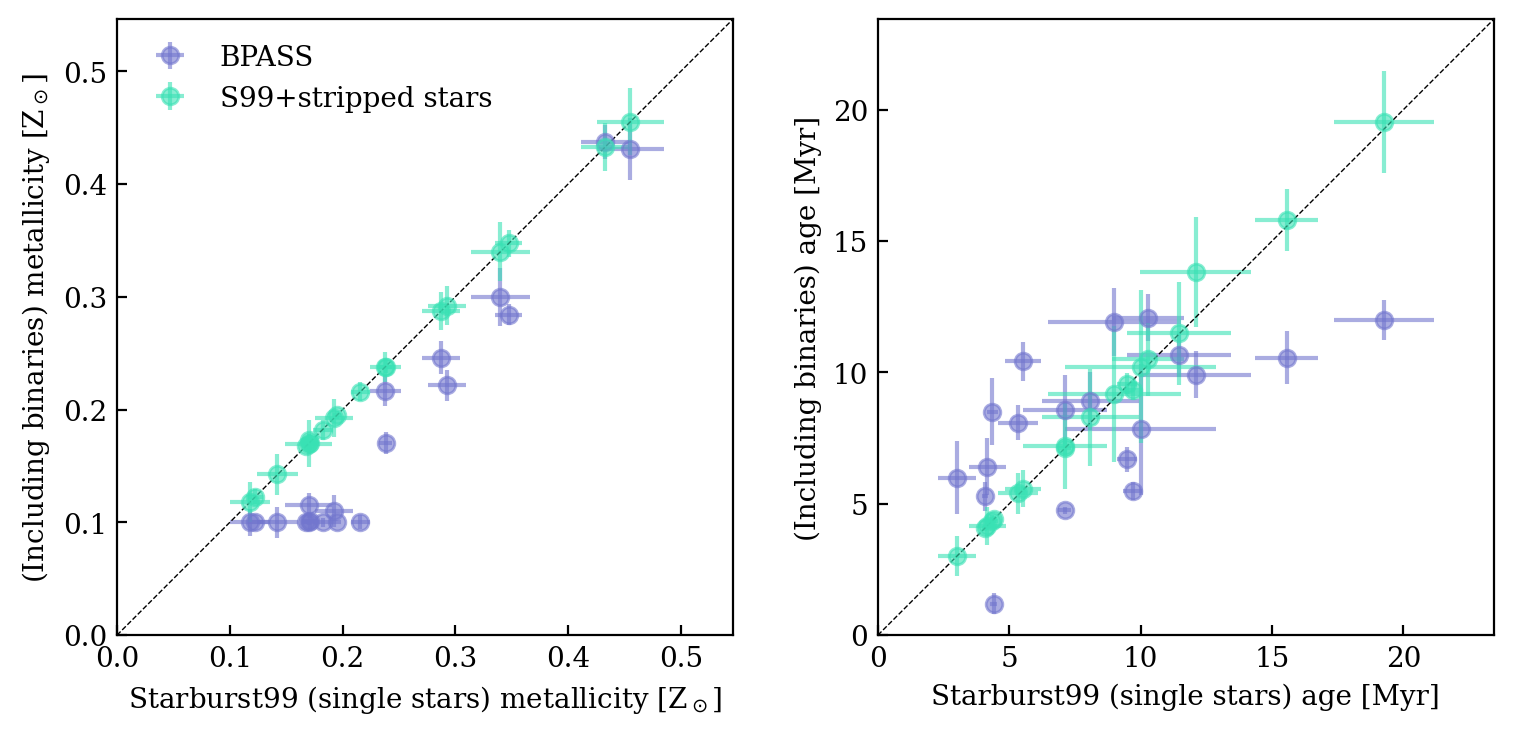}
    \caption{Comparison of metallicity (left) and age (right) for Starburst99 continuum fits compared with BPASS (purple) and Starburst99 with stripped stars (green). While the BPASS results generally agree with those from Starburst99, the stripped star results are virtually identical.}
    \label{fig:continuum}
\end{figure}

From FiCUS, we derive a light-weighted population age, metallicity, and E(B-V).
The two implementations of binaries yield different best-fit values for metallicity and age (Figure \ref{fig:continuum}). Including binaries using BPASS gives similar results to Starburst99, though overall slightly lower metallicities and some scatter in both properties, consistent with \cite{Chisholm2019}.

However, accounting for binaries by adding the \citetalias{Gotberg2019} stripped star models yields almost identical properties to Starburst99 alone. This similarity is because, while the stripped star models alone exhibit strong wind features for bursts of ages $\sim 10-25$ Myr, the overall magnitude of their spectra in the non-ionizing UV is a factor of $\sim 100$ less than that of the Starburst99 spectra alone. When added together, the stripped star wind features get washed out, and the overall contribution of stripped stars to the non-ionizing UV continuum is negligible.
The \citetalias{Gotberg2019} models, therefore, predict that stripped stars may be able to harden a population's ionizing flux without significant changes to the non-ionizing UV continuum.

\subsection{Predicted emission line fluxes}\label{subsec:lines}

\begin{figure}
    \centering
    \includegraphics[width=0.7\linewidth]{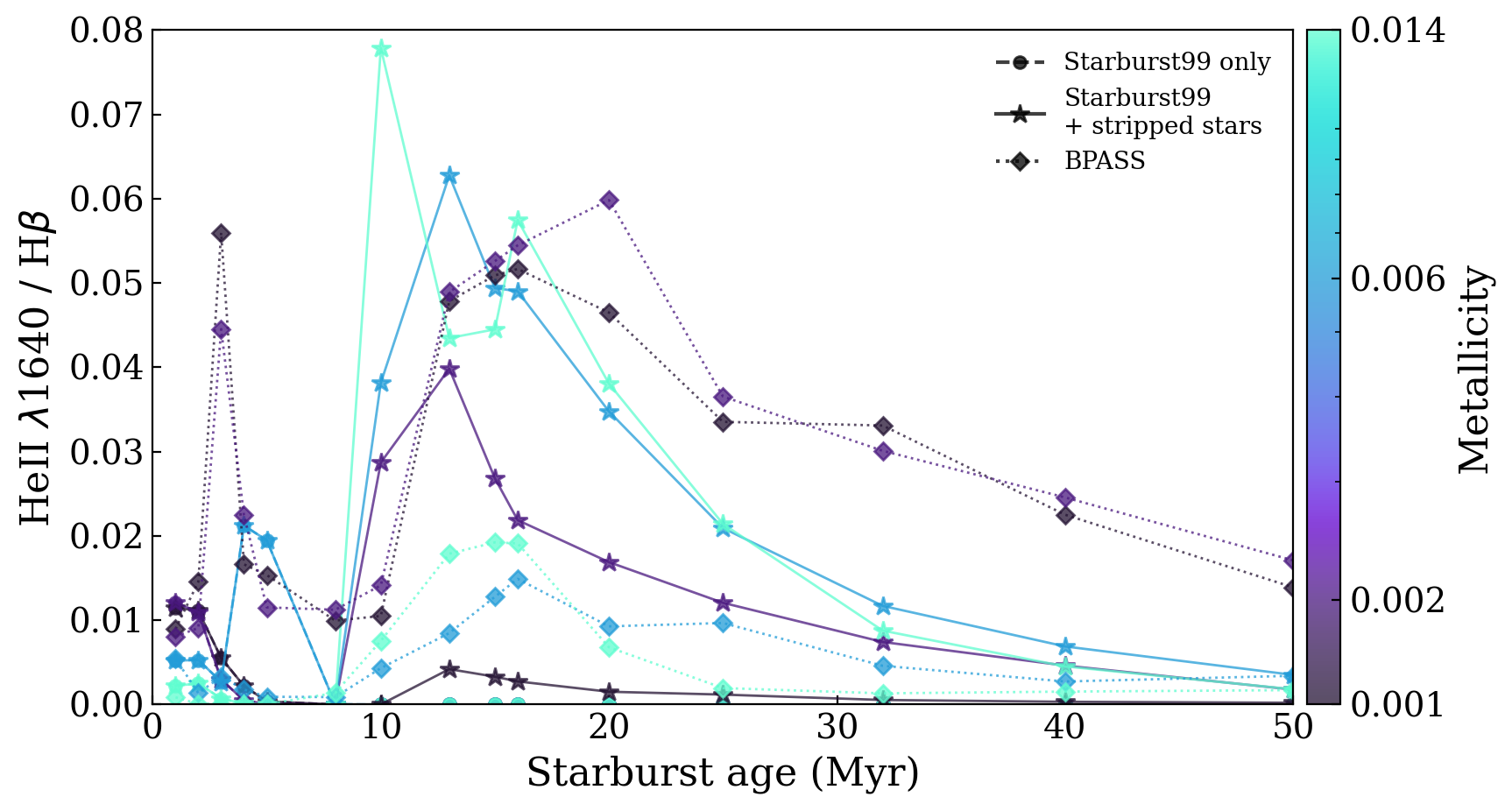}
    \caption{\heii/\hb\ as a function of time after a starburst, from Cloudy runs using incident spectra from Starburst99 (circles, dashed line), Starburst99 plus \citetalias{Gotberg2019} stripped stars (stars, solid line), and BPASS (diamonds, dotted line). The four metallicities in the model grid are indicated by color. Both BPASS and stripped stars boost \heii/\hb\ significantly beginning at 10 Myr.}
    \label{fig:heii}
\end{figure}

In addition to providing the nebular continuum, photoionization modeling with Cloudy yields predictions for nebular line emission. In Figure \ref{fig:heii}, we show how \heii/\hb \ changes over the life of a starburst, for all four metallicities and all three model grids.

For the Starburst99 models with single stars, this line ratio is strongest at lower metallicity and significant only at earlier times, becoming negligible by $\sim 5$ Myr.
The primary source of the nebular \heii \ emission line in these models is from massive Wolf-Rayet stars, which die out at these early times.
With the addition of the \citetalias{Gotberg2019} stripped star models starting at 10 Myr (stars, solid line), the line ratio value increases significantly, reaching values $\sim 3-30$ times higher than the Wolf-Rayet stars did at their peak. 

The BPASS models (diamonds, dotted line) show an overall similar trend in \heii/\hb, with initially high values due to emission from Wolf-Rayet stars that decrease as they die out, followed by even higher values at times $\geq 10$ Myr due to older binary products that produce hard ionizing radiation at later times.
There is one significant difference between the two implementations of binaries: for bursts with ages $\geq 10$ Myr, the highest \heii/\hb \ values are produced by the BPASS models with the \textit{lowest} metallicities but by the \citetalias{Gotberg2019} models with the \textit{highest} metallicities.
This may be another predicted consequence of the helium star desert at low metallicity in the \citetalias{Gotberg2019} models---a dearth of stripped stars should decrease the hardness of the ionizing spectrum.
However, we note that the stripped star atmosphere models used by \citetalias{Gotberg2019}, first presented by \cite{Gotberg2018}, were made prior to the first observations of intermediate mass stripped stars.
Their recent discovery has revealed weaker winds, which should affect their temperatures and, in turn, impact the \heii \ line.

The difference between BPASS and \citetalias{Gotberg2019} in Figure \ref{fig:heii} may also be related to the sources predominantly producing their ionizing radiation.
While the \citetalias{Gotberg2019} models contain only stripped stars, BPASS considers a range of other binary and stellar physics, including quasi-chemically homogeneous evolution (QHE). Though this phenomenon has not been conclusively observed, BPASS assumes that QHE can occur for stars at metallicities $\leq 0.004$ and masses $> 20 \msun$ \citep{Eldridge2017}. This metallicity cut-off corresponds with the jump in \heii/\hb \ between the BPASS models with $Z = 0.014,\ 0.006$ and $Z = 0.002,\ 0.001$, suggesting that QHE stars may be the most significant source of \heii \ photons in BPASS at later times.

\section{Conclusions and future work}

Different implementations of binaries in models have a significant impact on the predicted ionizing spectrum, and in turn, on the conclusions drawn from comparisons with data.
Because it requires photons with energies in excess of 54.4 eV, the nebular \heii\ emission line is sensitive to hard ionizing sources such as stripped stars, which can harden a population’s ionizing flux at later times after a starburst without significant changes to the non-ionizing spectrum.
We can use this line as a diagnostic for discriminating between models and studying binary populations, including testing whether a ``helium star desert'' exists at low metallicity.

We intend to continue this work by comparing observed and predicted \heii\ line fluxes. The observed emission line fluxes will be measured by subtracting the FiCUS best fit continuum from the observed spectrum and fitting the residual flux. The predicted emission line fluxes will be calculated by adding the emission line values output by Cloudy with the same weights from the FiCUS best fit.
By comparing the observed and predicted line fluxes, we can assess which models are able to self-consistently reproduce high-$z$ galaxies' spectra.

\bibliography{references}{}
\bibliographystyle{aanda}

\end{document}